\documentclass[twocolumn]{aastex631}

\usepackage{amsmath}
\usepackage{graphicx}
\def\bea{\begin{eqnarray}}
\def\eea{\end{eqnarray}}
\def\f{\frac}
\def\n{\nonumber}
\usepackage{threeparttable}
\usepackage{appendix}
\usepackage{chngcntr}
\usepackage{makecell}

\begin{document}

\title{Jet-Structure Imprint on the Curvature Tail of Gamma-Ray Burst Prompt Emission}
\correspondingauthor{Xiao-Hong Zhao, Hendrik J. van Eerten, Bin-Bin Zhang}
\email{zhaoxh@ynao.ac.cn}
\email{hjve20@bath.ac.uk}
\email{bbzhang@nju.edu.cn}

\author[0009-0008-2841-3065]{Zhen-Yu Yan}
\affiliation{School of Astronomy and Space Science, Nanjing University, Nanjing 210093, People’s Republic of China}
\affiliation{Key Laboratory of Modern Astronomy and Astrophysics (Nanjing University), Ministry of Education, People’s Republic of China}

\author[0000-0003-3659-4800]{Xiao-Hong Zhao}
\affiliation{Yunnan Observatories, Chinese Academy of Sciences, Kunming, People’s Republic of China}
\affiliation{Center for Astronomical Mega-Science, Chinese Academy of Sciences, Beijing, People’s Republic of China}

\author[0000-0002-8680-8718]{Hendrik J. van Eerten}
\affiliation{Department of Physics, University of Bath, Claverton Down, Bath BA2 7AY, UK}

\author[0000-0002-5485-5042]{Jun Yang}
\affiliation{Institute for Astrophysics, School of Physics, Zhengzhou University, Zhengzhou 450001, People’s Republic of China}

\author[0009-0002-3780-892X]{Jiang-Chuan Tuo}
\affiliation{School of Astronomy and Space Science, Nanjing University, Nanjing 210093, People’s Republic of China}
\affiliation{Key Laboratory of Modern Astronomy and Astrophysics (Nanjing University), Ministry of Education, People’s Republic of China}

\author[0000-0001-7599-0174]{Shu-Xu Yi}
\affiliation{Key Laboratory of Particle Astrophysics, Institute of High Energy Physics, Chinese Academy of Sciences, Beijing 100049, People’s Republic of China}

\author[0009-0008-8053-2985]{Chen-Wei Wang}
\affiliation{Key Laboratory of Particle Astrophysics, Institute of High Energy Physics, Chinese Academy of Sciences, Beijing 100049, People’s Republic of China}
\affiliation{University of Chinese Academy of Sciences, Chinese Academy of Sciences, Beijing 100049, People’s Republic of China}

\author[0009-0006-5506-5970]{Wen-Jun Tan}
\affiliation{Key Laboratory of Particle Astrophysics, Institute of High Energy Physics, Chinese Academy of Sciences, Beijing 100049, People’s Republic of China}
\affiliation{University of Chinese Academy of Sciences, Chinese Academy of Sciences, Beijing 100049, People’s Republic of China}

\author[0000-0002-4771-7653]{Shao-Lin Xiong}
\affiliation{Key Laboratory of Particle Astrophysics, Institute of High Energy Physics, Chinese Academy of Sciences, Beijing 100049, People’s Republic of China}

\author[0000-0003-4111-5958]{Bin-Bin Zhang}
\affiliation{School of Astronomy and Space Science, Nanjing University, Nanjing 210093, People’s Republic of China}
\affiliation{Key Laboratory of Modern Astronomy and Astrophysics (Nanjing University), Ministry of Education, People’s Republic of China}

\begin{abstract}
Even though the prompt emission of gamma-ray bursts (GRBs) is highly beamed, high-latitude emission still produces a distinct light curve break after the intrinsic emission ceases and the edge of the jet comes into view. This curvature effect offers a direct probe of the jet structure during the prompt phase. To uncover the geometric structure of the GRB jet encoded in the prompt light-curve evolution, we develop a numerical model that calculates synchrotron light curves from structured jets to interpret the observed break. We apply this model to the prompt emission of GRB 230307A, which displays a rare late-time break. Our analysis demonstrates that simple spherical outflow and top-hat jet models are inadequate to reproduce the light curve. Instead, the observations are best described by a power-law wing jet with a uniform core ($\theta_{\rm core}=0.0147$ rad) and a surrounding power-law wing. Our results demonstrate that the break in late-time prompt emission can be a powerful diagnostic of GRB jet structure.
\end{abstract}


\section{Introduction} \label{sec:Introduction}
Over the past decades, thanks to improvements in detector sensitivity, the number of gamma-ray burst (GRB) detections has significantly increased \citep{2003A&A...411L...1W,2004ApJ...611.1005G,2005SSRv..120..143B,2009ApJ...697.1071A,2009ApJ...702..791M,2013ApJS..207...38P,2022RDTM....6...12L,2023NIMPA105668586Z}. These observations provide strong evidence that GRBs originate from relativistic jets \citep{1999ApJ...525..737R,1999ApJ...519L..17S,1998ApJ...499..301M}. Early models often assumed these jets to be simple, uniform “top-hat” outflows \citep{1997ApJ...482L..29M,1997MNRAS.288L..51W,1997ApJ...487L...1R,2008MNRAS.390..675R}. However, growing observational and theoretical evidence reveals that GRB jets possess more complex angular structure \citep{2002MNRAS.332..945R,2002ApJ...571..876Z,2003ApJ...591.1075K,2014MNRAS.440.3292L,ghirlandaCompactRadioEmission2019}. This wealth of high-quality data necessitates the development of more sophisticated theoretical models to connect jet physics to observable signals. This raises two fundamental questions: (1) how do different jet structures (e.g., angular variations in energy and Lorentz factor) shape the observed light curves and spectra? (2) how reliably can these observables constrain the jet's structure? 

To understand how jet profiles shape GRB emission, we must model the observable signatures of different structures. The simplest and most widely used model is the top-hat jet, which assumes a uniform conical outflow with sharp edges. Beyond this, more complex structures have been proposed, such as power-law jets \citep{1998ApJ...499..301M,2001ApJ...552...72D,2002MNRAS.332..945R,2002ApJ...571..876Z,2003ApJ...591.1086G}, Gaussian jets \citep{2002ApJ...571..876Z,2003ApJ...591.1075K,2004ApJ...601L.119Z,2004ApJ...601..371L,2005ApJ...621..875D}, two-component jets \citep{2003Natur.426..154B,2004ApJ...605..300H,2005ApJ...626..966P,2005MNRAS.357.1197W,2008Natur.455..183R}, and ``mini-jets" in broad jets \citep{2000ApJ...535..152K,2009MNRAS.394L.117N,2011ApJ...726...90Z,2015ApJ...805..163D,2025ApJ...985..239Y}. Theoretically, these jet geometries should leave distinct signatures on the GRB afterglow, such as achromatic breaks and post-break decay indices in the light curve \citep{2000ApJ...541L...9K,2001ApJ...562L..55F,2001Natur.414..853W,2012ApJ...751...57D,2016ApJ...825...97U,2019A&A...628A..59O}, offering a potentially powerful means to distinguish between models. In practice, however, these afterglow-based constraints on jet structure remain highly uncertain because parameter degeneracies involving the circumburst density profile, microphysical parameters, jet opening angle, and viewing angle often prevent a robust determination of the structure.

If prompt emission can be used to constrain jet structure, it would provide an independent probe to break these degeneracies and cross-validate afterglow interpretations. However, inferring jet structure directly from the prompt phase faces a significant challenge. During this phase, the relativistic beaming effect concentrates the observed radiation to a very narrow angle, $\theta< 1/\Gamma$, where $\Gamma$ is the bulk Lorentz factor of the jet. As long as the jet's half-opening angle, $\theta_{\rm jet}$, is larger than this relativistic cone ($\theta_{\rm jet}>1/\Gamma$), the high-latitude emission will be swamped by emission from closer to the line-of-sight to the observer, if not too weak to be detectable altogether. Consequently, the emission is insensitive to the jet's overall angular profile, making it nearly impossible to distinguish a structured jet from a much wider uniform outflow or even a spherical fireball.

However, for sufficiently bright bursts, the jet structure can be revealed at late times by the ``curvature effect" \citep{1996ApJ...473..998F,2000ApJ...541L..51K,2004ApJ...614..284D,2005astro.ph.11699D,2009ApJ...703.1696Z,2015ApJ...808...33U}. After the central engine ceases activity or the injection of high-energy electrons ends, the observed flux is dominated by delayed photons from progressively higher latitudes on the curved emitting surface, where $\theta$ eventually exceeds $1/\Gamma$. If the jet has a finite edge at a narrow angle $\theta_{\rm jet}$, the absence of photons from beyond this boundary will produce a sharp and achromatic break in the light curve, directly revealing the jet’s opening angle. We identify this behavior as the “curvature-tail effect”, which serves as an observational probe of the jet geometry.

This pure geometric effect is best understood in terms of the equal-arrival-time surface (EATS) \citep{1997ApJ...491L..19W,1999ApJ...513..679G,2007ChJAA...7..397H,2015ApJ...808...33U}: following the onset of prompt emission arriving from the jet tip, the EATS expands to larger latitude. As it approaches the jet boundary, the observed decay becomes directly sensitive to whether the jet edge is sharp or smoothly structured. Once it crosses $\theta_{\rm jet}$, the missing emission causes the light curve to steepen. For a simple top-hat jet, the flux decays sharply, while if additional emission arises from angular structure beyond $\theta_{\rm jet}$, as in structured jet models, the decay will inevitably be shallower. This contrast in the post-break behavior, therefore, provides a powerful diagnostic to distinguish different jet structures, free from the uncertain external medium profile and microphysical parameters that often affect the afterglow-based constraints.

The ideal candidates for using these observable signatures to constrain jet structure are highly luminous GRBs with narrow jets, where geometric effects such as the curvature-tail feature are pronounced enough for detection. Such signatures are more likely to be detected when the prompt emission is well covered at lower energies, such as in the X-ray band. Modeling these features in the prompt emission allows us to directly probe the jet structure, providing an independent constraint that helps break parameter degeneracies often encountered in afterglow studies. The recent bright burst GRB 230307A, which exhibits a rare late-time break in its prompt emission \citep{2025NSRev..12E.401S}, presents a perfect test case for this methodology, motivating the comprehensive and quantitative study of the curvature-tail effect presented in this work.

This paper is structured as follows. In Section \ref{sec:curvature-tail Effect}, we provide a detailed geometric explanation of the curvature-tail effect. In Section \ref{sec:MODEL}, we describe the synchrotron radiation model implemented in our numerical code and predict what observable properties the curvature-tail effect can cause under different jet structures with canonical physical parameters. We then apply this framework in Section \ref{sec:TheFit}, where we fit the late-time light curve of GRB 230307A using different jet structures. Finally, we present a broader discussion of our results in Section \ref{sec:Discussion}.

\section{Physical picture of the curvature-tail effect} \label{sec:curvature-tail Effect}

The curvature-tail effect discussed in this paper differs in its physical cause between the prompt and afterglow phases, although both originate from the same geometric effect associated with the finite angular extent of the jet. In the afterglow stage, the so-called jet break arises when the jet decelerates and the relativistic beaming angle expands to $1/\Gamma \sim \theta_{\rm jet}$, causing the observer to lose radiation from outside the jet boundary \citep{1997ApJ...487L...1R,1999MNRAS.306L..39M,1999ApJ...526..707P,1999ApJ...519L..17S,1999ApJ...523L.121H}. In contrast, the curvature-tail effect studied here occurs in the prompt emission phase, when the intrinsic radiation ceases abruptly and the observer begins to receive photons from progressively higher latitudes ($>1/\Gamma$) until the jet boundary is reached. The resulting rapid decay or break in the light curve is therefore caused by the cessation of radiation rather than by the jet’s deceleration.

To show the physical picture of the curvature-tail effect, a detailed geometric system of the jet will be established in this section. Consider the simplest case of a uniform “top-hat’’ jet in GRBs, moving at a highly relativistic constant speed with bulk Lorentz factor $\Gamma$. The jet has a half-opening angle $\theta_{\rm jet}$, and the line of sight is along the central axis of the jet. The total observed flux from the jet is calculated by integrating the emission over the visible portion of the outflow. Definitions of key times are listed in Table \ref{table:time}.

At any given moment, the photons arriving at the observer originate from different locations and times within the jet, defining a geometry known as the EATS. For a relativistic outflow, there is a precise geometric relationship among the emission radius ($R$), the angle position ($\theta$) on the EATS, and the observer's time after the trigger of the burst ($t_{\rm obs}$). $\theta$ is the angle between the line of sight and the orientation from the central source to a point on the EATS. The inner edge radius of the emission region is $R_0$ and the emission radius $R$ ($R\ge R_0$) on the EATS is here defined as:
\begin{equation} \label{eq:EATS}
R=\frac{\beta c (t_{\rm obs,0} + t_{\rm obs})}{(1+z)(1-\beta \cos\theta)},
\end{equation}
where $\beta$ is the dimensionless speed of the jet, $c$ is the speed of light, $t_{\rm obs,0}=R_0(1-\beta)(1+z)/(\beta c )$, $z$ is redshift. From Equation \eqref{eq:EATS}, the angle position can be derived as: 
\begin{equation} \label{eq:theta}
\theta=\arccos{\left[ \frac{1}{\beta} - \frac{c (t_{\rm obs,0} + t_{\rm obs})}{(1+z)R} \right] }.
\end{equation}
Equation \eqref{eq:theta} defines the EATS as a surface in ($R$, $\theta$) space for a given $t_{\rm obs}$. An analysis of this relation shows that for a certain $t_{\rm obs}$, the maximum angle, $\theta_{\rm max}(t_{\rm obs})$, corresponds to the minimum possible emission radius, $R_{\rm min}$. For a shell that starts injecting high-energy electrons and emitting photons from $R_{\rm 0}$, this minimum radius is $R_{\rm min}=R_{\rm 0}$, and the maximum angle of the EATS contributing to the observed emission at this radius is:
\begin{equation} \label{eq:theta_max}
\theta_{\rm max}(t_{\rm obs})=\theta_{R_0}(t_{\rm obs})=\arccos{\left[ 1-\frac{ct_{\rm obs}}{(1+z)R_0}\right]}. 
\end{equation}

Equations \eqref{eq:theta} and \eqref{eq:theta_max} demonstrate that the maximum angle subtended by the EATS within the emission region, $\theta_{\rm max}(t_{\rm obs})$, is a monotonically increasing function of observer time, $t_{\rm obs}$. If this angle is smaller than jet half-opening angle within burst duration, $\theta_{\rm max}(t_{\rm obs})<\theta_{\rm jet}$, the EATS remains entirely within the jet cone throughout the emission region, so that no curvature-tail signature is expected in the prompt light curve. However, for a GRB with sufficiently narrow jet, long duration, and small emission radius, the EATS can exceed the jet boundary already early on, $\theta_{\rm jet}<\theta_{\rm max}(t_{\rm obs})$, leading to a break in the light curve that is possible to be detected during the prompt stage. Moreover, $\theta_{\rm max}(t_{\rm obs})$ is independent of $\Gamma$. Therefore, the magnitude of $\Gamma$ does not affect the detection of the curvature-tail effect.

\begin{figure}[ht!]
\centering
\includegraphics[width=0.96\linewidth]{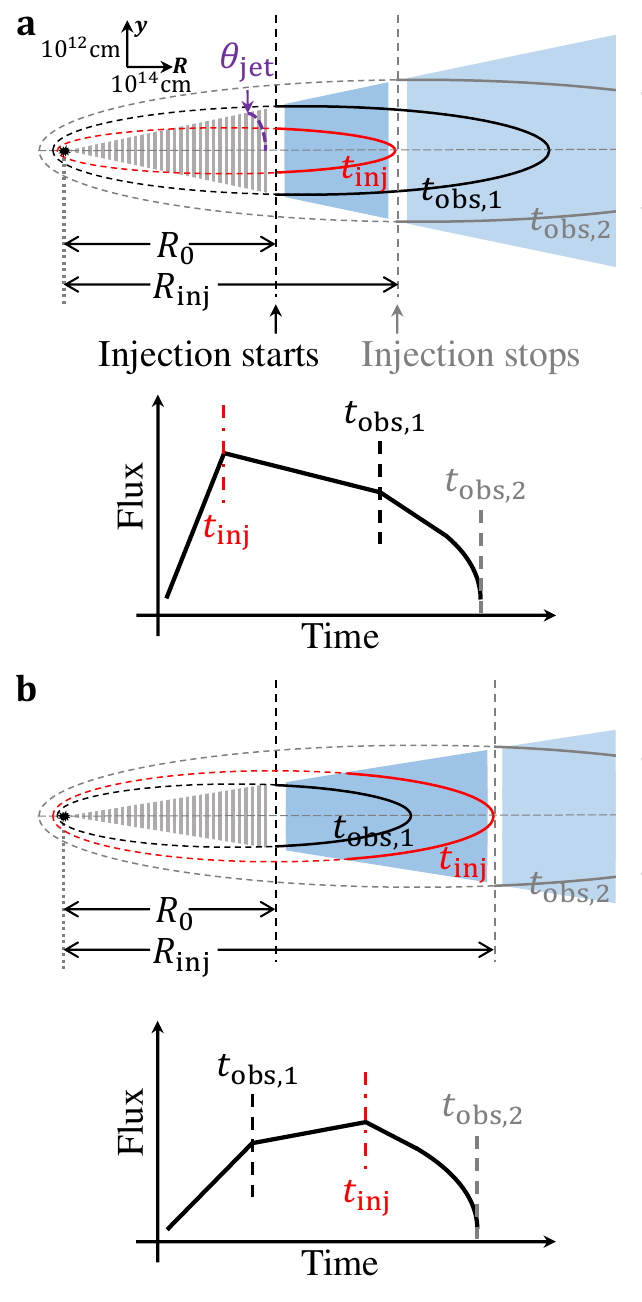}
\caption{ Curvature-tail effect diagram. The panels a and b show two scenarios where the curvature-tail effect occurs in decaying and rising phases, respectively. The cone shows the history of a jet sweeping trajectory through the outside space. The ellipse represents EATS, and its solid section contributes to the observed flux. The cone in the grey shaded region indicates the fraction that does not contribute to the observed flux, while the cone in dark or light blue does. The dark blue region represents the electron injection region, while the light blue region has no injection. The horizontal black dashed line represents the line of sight for an on-axis observer. The purple angle represents the jet half-opening angle. The vertical black and grey dashed lines in the jet diagram represent the radius where the injection starts and stops, respectively. 
\label{fig:curvature-tail}}
\end{figure}

Figure \ref{fig:curvature-tail} illustrates the geometric origin of the curvature-tail effect and how these breaks arise. The cone represents the trajectory of a thin-shell jet propagating outward from the central engine. The portions of the jet within $\theta\leq\theta_{\rm jet}$ in dark blue indicate the parts of the jet that can contribute to the observed emission. At any given observed time, only the section of the EATS lying inside this blue region contributes to the observed flux, which is plotted in the solid ellipse shape. At early times, $\theta_{\rm max}(t_{\rm obs})<\theta_{\rm jet}$, and the EATS lies entirely inside the jet cone. As observed time increases, the EATS expands outward. When the observer time reaches a critical time of $t_{\rm obs}=t_{\rm obs,1}$, the maximum angle on the EATS equals the jet's half-opening angle, i.e, $\theta_{\rm max}(t_{\rm obs,1})=\theta_{\rm jet}$. This critical time is defined by :
\begin{equation} \label{eq:theta_jet}
\theta_{\rm max}(t_{\rm obs,1})=\arccos{\left[ 1-\frac{c t_{\rm obs,1}}{(1+z)R_0}\right]} \equiv\theta_{\rm jet}. 
\end{equation}
Then $t_{\rm obs,1}$ can be obtained by
\begin{equation} \label{eq:tobs1}
t_{\rm obs,1}=\frac{R_0(1+z)(1- \cos\theta_{\rm jet})}{c }.
\end{equation}
Beyond this time, the EATS continues to expand, but the portion at $\theta_{\rm jet}<\theta<\theta_{\rm max}(t_{\rm obs})$ lies outside the jet and thus ceases to contribute. The resulting loss of high-latitude emission causes the observed flux to decay more rapidly, creating a characteristic break in the light curve. After some time, the EATS eventually touches the jet edge at the injection stopping radius, $R_{\rm inj}$, at the time of $t_{\rm obs,2}$, which can be given by
\begin{eqnarray} \label{eq:tobsout}
t_{\rm obs,2}&&=(R_0+2\Gamma^2 \beta c t_{\rm inj}/(1+z))\times \nonumber\\ \quad &&\frac{(1+z)(1-\beta \cos\theta_{\rm jet})}{\beta c }-t_{\rm obs,0} \nonumber \\
&&=(t_{\rm obs,0}+t_{\rm inj})\frac{1-\beta \cos\theta_{\rm jet}}{1-\beta }-t_{\rm obs,0},
\end{eqnarray}
where $t_{\rm inj}$ is the total injection time in the observer frame, and the electron injection region is limited in the area between $R \in [R_0,~R_{\rm inj}]=[R_0,~R_0+2\Gamma^2 \beta c t_{\rm inj}/(1+z)]$. After that, all emissions are from the cooling low-energy electrons, causing a much steeper decay turning into a cut-off.

For the fast-cooling regime, this injection stopping radius, $R_{\rm inj}$, can be approximately regarded as the emission stopping radius. Moreover, $t_{\rm inj}$ can be regarded as the peak time of the overall light curve. If this time is smaller than the curvature-tail effect starting time, $t_{\rm inj}<t_{\rm obs,1}$, the curvature-tail effect steepens the late-time light curve and can cause a break as shown in panel a of Figure \ref{fig:curvature-tail}. If it is the opposite situation, $t_{\rm obs,1}<t_{\rm inj}$, the curvature-tail effect can even occur in the rising phase and cause a slower rise as shown in panel b of Figure \ref{fig:curvature-tail}. The latter situation requires an extremely long injection duration and will have a shorter decay duration right after the peak time. These requirements make the rise time at least comparable to, or even longer than, the decay time, which differs from the common fast-rise and exponential-decay \citep{1996ApJ...473..998F} shape of GRB light curves that we often observe. Without loss of generality, our study therefore focuses on the former situation, $t_{\rm inj}<t_{\rm obs,1}$.

\section{Quantitative modeling of the curvature-tail effect signatures} \label{sec:MODEL}
\subsection{synchrotron emission model} \label{sec:Synchrotron Radiation Model}
To quantitatively compute the observable signatures of the curvature-tail effect from the flux of relativistic jets, a standard synchrotron emission model is adopted. The model used in this analysis is an extension of the numerical framework we developed in \cite{2023ApJ...947L..11Y,2024ApJ...962...85Y}. In that work, we modeled synchrotron emission from an expanding thin shell with a constant Lorentz factor in a decaying magnetic field. This model was successful in jointly fitting all the time-resolved spectra of the GRBs with a single set of physical parameters. In this work, we have enhanced this framework to incorporate more complex jet structures. This subsection provides a brief overview of the model's definitions and details the new geometric components and physical parameterizations.

The physical model we consider in this work is a thin shell with a relativistic speed, containing injected electrons that radiate synchrotron emission in a magnetic field. In this model, the flux $F_{\nu_{\rm obs}}$ at frequency $\nu_{\rm obs}$ observed by an on-axis observer can be expressed as \citep{2024ApJ...962...85Y}
\begin{eqnarray} \label{eq:Fv_total}
&&F_{\nu_{\rm obs}} = F_{\nu_{\rm obs}}(t_{\rm obs},\nu_{\rm obs}, B^{\prime}_{\rm 0}, \alpha_B, \gamma^{\prime}_{\rm m}, \Gamma, p, t_{\rm inj}, R_{\rm 0}, Q_{\rm 0}) \nonumber \\
&& =C_0 \frac{ B^{\prime}_{\rm 0} {R_{\rm 0}}^{\alpha_B}} {c^{\alpha_B }{\Gamma}^3 \beta^{\alpha_B +1}} \times \\
&& \int_{t_{\rm e,0}}^{t_{\rm e,max}} 
 \frac{{t_{\rm e}}^{-\alpha_B-1}} {{\left( 1-\beta \cos\theta\right)}^2} 
 \int_{\gamma^{\prime}_{\rm e,min}}^{\gamma^{\prime}_{\rm e,max}} 
 \left(\frac{{\rm d} N_{\rm e}^{\prime}}{{\rm d} \gamma^{\prime}_{\rm e}}\right) 
 F\left(\frac{\nu^{\prime} }{\nu_{\rm c}^{\prime}}\right) {\rm d} \gamma^{\prime}_{\rm e} \;\; {\rm d} t_{\rm e}, \nonumber 
\end{eqnarray}
where $B'_0$ is the initial magnetic field strength at $R_0$, $\alpha_B$ is the magnetic field decaying power-law index, $\gamma_{\rm m}^{\prime}$ is the minimum injection energy of electron, $p$ is the initially injected electron distribution's the power-law index, $Q_0$ is an electron injection factor in ${\rm cm}^{-1}$, $C_0=\sqrt{3} {q_{\rm e}}^3 (1+z)/ ({8\pi m_{\rm e} c^{2} {D_{\rm L}}^2})$, $q_{\rm e}$ is electron charge, $D_{\rm L}$ is luminosity distance, $t_{\rm e}$ is the time variable in the source frame, $t_{\rm e,0}$ and $t_{\rm e,max}$ are the initial time and the maximum time in the source frame, $\gamma^{\prime}_{\rm e,min}$ and $\gamma^{\prime}_{\rm e,max}$ are the minimum and maximum electron energy, ${\rm d}N_{\rm e}^{\prime}/{\rm d}\gamma^{\prime}_{\rm e}$ is electron distribution, $F(x)=x \int_{x}^{+\infty} K_{\rm 5 / 3}(k) d k$, and $ K_{\rm 5 / 3}(k)$ is the Bessel function. We emphasize that $t_{\rm e}$ is defined for a given shell and set to be proportional to $R$, $R=c\beta t_{\rm e}$.

\subsection{the curvature-tail effect based on the top-hat jet model} \label{sec:Top-hat jet model}
Using the synchrotron emission model, we will show the curvature-tail effect considering the simplest case of a uniform “top-hat” jet. In our previous works \citep{2023ApJ...947L..11Y,2024ApJ...962...85Y}, the jet structure was not considered, which means the shell was regarded as an isotropic spherical shell. Therefore, the integration area of the EATS at certain observer time is from the initial emission radius to the maximum radius of the EATS. The value of the initial emission radius depends on the specific energy dissipation mechanism, for example, the collision of internal shocks or the reconnection of magnetic fields. The maximum radius on the EATS at certain observer time is along the direction of the line of sight. These determine the definitions of the lower and upper time limits, $t_{\rm e,0}=R_0/\beta c$ and $t_{\rm e,max}=t_{\rm e,0}+t_{\rm obs}/(1+z)/(1-\beta)$ in Equation \eqref{eq:Fv_total}. However, once the jet nature of the outflow is considered in the calculation, the curvature-tail effect mentioned in Section \ref{sec:curvature-tail Effect} should also be considered, and the integration shown in Equation \eqref{eq:Fv_total} should be modified. For a top-hat jet, the presence of a sharp edge at the opening angle truncates the domain of integration. More precisely, the integral for the observed flux in Equation \eqref{eq:Fv_total} is no longer performed over all the time (angle), but is instead limited to a smaller range with a higher lower time limit:
\begin{equation} \label{eq:lower time limit}
t_{\rm e,0}=
\begin{cases}
R_0/\beta c, & t_{\rm obs}<t_{\rm obs,1}\\
\dfrac{ t_{\rm obs,0}+t_{\rm obs}}{(1+z)(1-\beta \cos\theta_{\rm jet})}, & t_{\rm obs} \ge t_{\rm obs,1}
\end{cases},
\end{equation}
where the line of sight is assumed to be along the central line of the jet cone, and we maintained the assumption of this on-axis observation in this work.

\begin{figure}[ht!]
\centering
\includegraphics[width=1.0\linewidth]{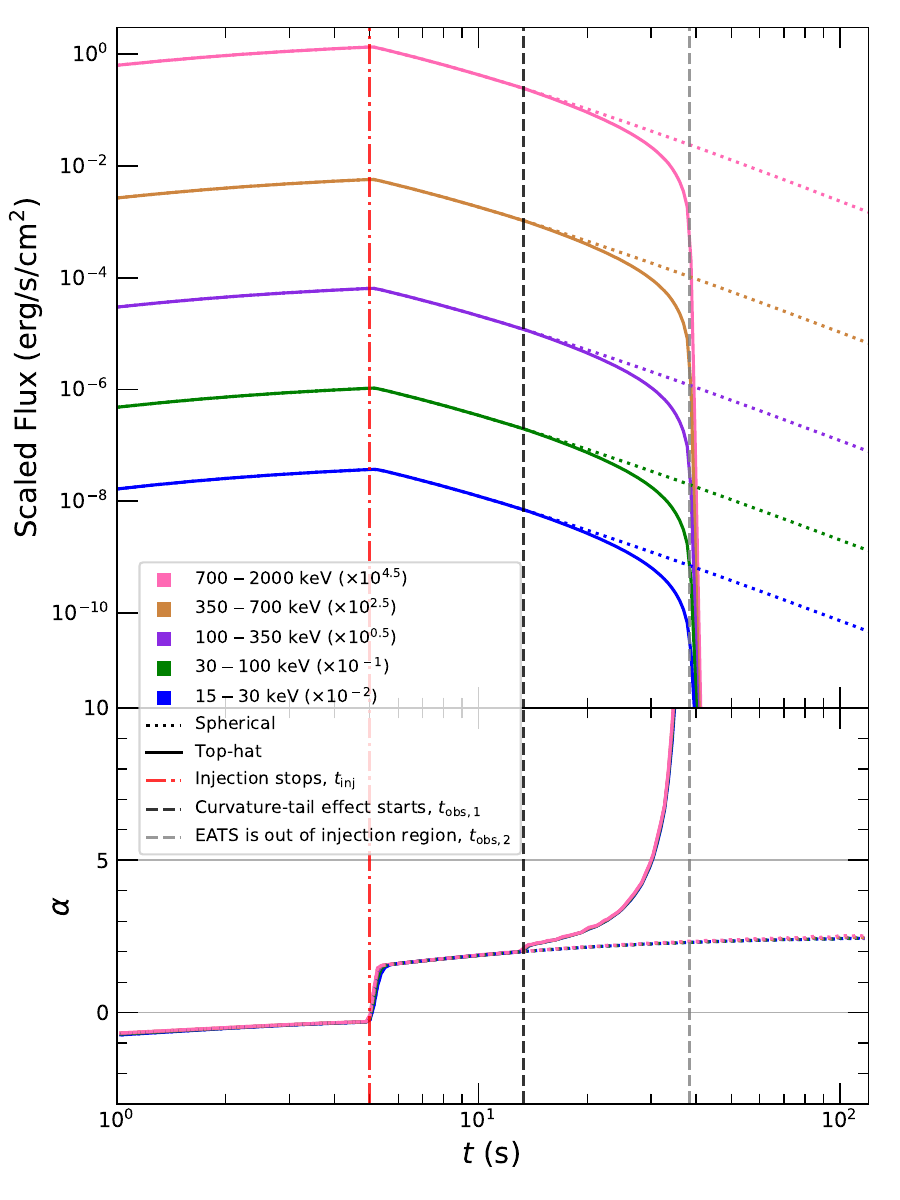}
\caption{Scaled flux on a log-log scale and temporal decay indices. The flux is calculated using the canonical physical parameters. Colored dotted and solid curves represent the spherical model and the top-hat jet model ($\theta_{\rm jet}=0.02~{\rm rad}$), respectively. Three key times in Figure \ref{fig:curvature-tail} are also plotted as vertical lines.
\label{fig:model_gradually_tophat}}
\end{figure}

To present how the change in Equation \eqref{eq:lower time limit} affects the actual observed GRB signals, we adopt canonical physical parameters: $z=1$, $B'_0=10^{2}~{\rm G}$, $\alpha_B=0.5$, $\gamma^{\prime}_{\rm m}=10^{6}$, $\Gamma=10^{2}$, $p=2.5$, $t_{\rm inj}=5~{\rm s}$, $R_{\rm 0}=10^{15}~{\rm cm}$, $Q_{\rm 0}=10^{32}~{\rm cm^{-1}}$. Note that the new definition of $Q_0$ differs from the model of \cite{2023ApJ...947L..11Y,2024ApJ...962...85Y} in the injection term of the electron continuity equation during the injection time $t<t_{\rm inj}$, and the new form is
\begin{equation} \label{eq:injection}
Q'\left( \gamma^{\prime}_{\rm e}\right)=
\begin{cases}
Q_{\rm 0} {\left( \dfrac{\gamma^{\prime}_{\rm e}}{\gamma^{\prime}_{\rm m} }\right)}^{-p}, & \gamma ^{\prime} _{\rm m} < \gamma^{\prime}_{\rm e} < \gamma^{\prime}_{\rm e,max}\\
0, & \text{otherwise}
\end{cases}.
\end{equation}
We also choose five energy bands (15-30, 30-100, 100-350, 350-700, 700-2000 keV) to study the differences between the bands. We then use our model Equation \eqref{eq:Fv_total} to calculate the expected light curves for a spherical model and a top-hat jet model. The results are shown in Figure \ref{fig:model_gradually_tophat}. 

In the upper panel of Figure \ref{fig:model_gradually_tophat}, the observed scaled flux calculated by the spherical model is plotted in colored dotted curves. The colored solid curves represent the model with a top-hat jet where $\theta_{\rm jet}=0.02 ~{\rm rad}$. Three key times are demonstrated in the plot by three vertical lines: (1) the artificially defined time when electron injection stops plotted in red dash-dotted line, $t_{\rm inj}=5 ~ {\rm s}$; (2) the model-predicted time when the curvature-tail effect should start with given parameters plotted in black dashed line, $t_{\rm obs,1}=13.3 ~ {\rm s}$, obtained by Equation \eqref{eq:tobs1}; (3) the model-predicted time when the EATS is totally outside of electron injection region in grey dashed line, $t_{\rm obs,2}=38.3 ~ {\rm s}$, obtained by Equation \eqref{eq:tobsout}. The lower panel shows the corresponding temporal decay indices, $\alpha$, where $F_{\nu}\propto t_{\rm obs}^{- \alpha}$, derived from the light curves in the upper panel. These indices overlap with each other, revealing that the indices of different energy bands evolve similarly.

As shown in Figure \ref{fig:model_gradually_tophat}, the curvature-tail effect in the top-hat jet can theoretically cause a break to the light curves in the prompt emission phase under the conditions mentioned above. Around $t_{\rm obs,1}$, the flux light curves in solid curves have a steepening phenomenon, which is clearer in the lower panel. The lower panel shows that the temporal decay indices actually increase faster right after the curvature-tail effect starting time $t_{\rm obs,1}$. These changes reveal that the second derivatives of the flux on a log-log scale have discontinuous changes at $t_{\rm obs}=t_{\rm obs,1}$. These theoretically predicted features provide a direct diagnostic for observations: a curvature-tail break should appear as an achromatic steepening across energy bands, with the post-break index increasing faster than expected from simple radiative cooling. Identifying $t_{\rm obs,1}$ can help constrain the jet opening angle, offering a signature of the jet edge.

Moreover, the light curve of the top-hat jet model has a cut-off at $t_{\rm obs}=t_{\rm obs,2}$. This is because the total integration area of the EATS inside the jet cone is beyond the maximum electron injection radius, and no newly injected electrons are in the EATS. Therefore, all the observed photons are from old cooling electrons. Given the canonical physical parameters, those remaining electrons in the emission region will soon lose their energy in a fast cooling regime. This is also the reason why the light curves in all energy bands evolve similarly.

\subsection{the curvature-tail effect based on structured jet models} \label{sec:wing structure model}
As a comparison with the above top-hat jet model, we investigate three structured jet models in this subsection: (1) a two-component jet with a uniform core and a distinct uniform wing; (2) a jet with a uniform core and a power-law wing; and (3) a jet with a uniform core and a Gaussian wing. For each model, we will define its geometry and present its predicted theoretical light curve to study how these structures will change the curvature-tail effect.

The first model we consider is a two-component jet, comprising a uniform core surrounded by a uniform wing. We parameterize the structure by assuming that the energy injected into electrons differs between the two regions. Specifically, the injected electron energy spectrum is assumed to have the same power-law index in both the core and the wing, but with different normalizations. We define the energy injected in the wing to be a fraction, $f<1$, of the energy injected in the core, such that the wing is less energetic:
\begin{equation} \label{eq:injection_twocomponent}
Q'\left( \gamma^{\prime}_{\rm e}\right)=
\begin{cases}
Q_{\rm 0} {\left( \dfrac{\gamma^{\prime}_{\rm e}}{\gamma^{\prime}_{\rm m} }\right)}^{-p}, & 0 < \theta < \theta_{\rm core}\\
fQ_{\rm 0} {\left( \dfrac{\gamma^{\prime}_{\rm e}}{\gamma^{\prime}_{\rm m} }\right)}^{-p}, & \theta_{\rm core} < \theta < \theta_{\rm w}
\end{cases},
\end{equation}
where $\theta_{\rm w}=\theta_{\rm core}+\Delta\theta_{\rm wing}$, $\theta_{\rm w}$ is wing angle, and $\Delta\theta_{\rm wing}$ is wing width. In this subsection, this coefficient is fixed as $f=0.5$. We also assume that the constant bulk Lorentz factors of the two components are the same.

\begin{figure}[ht!]
\centering
\includegraphics[width=1.0\linewidth]{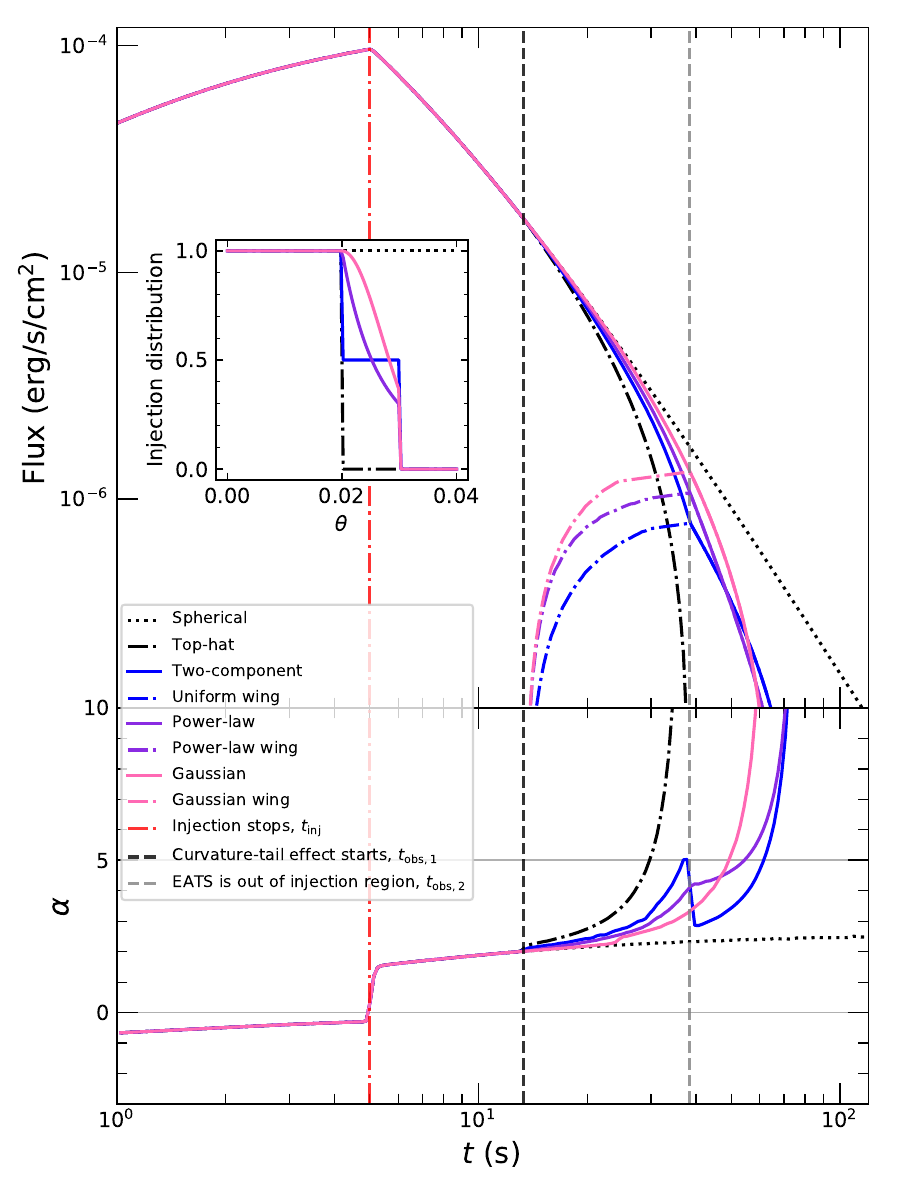}
\caption{Scaled flux in 15--2000 keV on a log-log scale and temporal decay indices of the jet models with wing structures. The nested subplot in the upper panel presents the normalized angular injection distribution. The flux is calculated using the canonical physical parameters. The blue, purple, and pink solid curves represent the two-component model, power-law wing model, and Gaussian wing model, respectively, which contain a core with $\theta_{\rm core}=0.02~{\rm rad}$ inside a wing with $\Delta\theta_{\rm wing}=0.01~{\rm rad}$. The black dotted curve represents the spherical model. The black dash-dotted curve represents the top-hat model (the same as the contribution of the first uniform core component). The colored dash-dotted curves represent the wings' independent contributions. Three key times in Figure \ref{fig:curvature-tail} are also plotted as vertical lines. 
\label{fig:model_gradually_two}}
\end{figure}

In this subsection for this two-component jet model, the wing width is fixed as $\Delta\theta_{\rm wing}=0.01 ~ {\rm rad}$, the jet core is fixed as $\theta_{\rm core}=0.02 ~ {\rm rad}$, and all the other parameters are the same as the above canonical physical parameters. The flux calculated in 15--2000 keV is presented in blue in Figure \ref{fig:model_gradually_two}. For the internal component, $\theta<\theta_{\rm core}=0.02 ~ {\rm rad}$, $t_{\rm obs,1}=13.3~ {\rm s}$, derived from Equation \eqref{eq:tobs1}, and $t_{\rm obs,2}=38.3~ {\rm s}$, derived from Equation \eqref{eq:tobsout}, which are the same as those of the top-hat jet model. Therefore, the internal component can be simply regarded as the top-hat model in the black dash-dotted curve. It is worth mentioning that here $t_{\rm obs,1}$ is not only the time when the curvature-tail effect of the core starts, but the time when the second component starts to contribute to the observed flux. The light curves are dominated by the two components successively. Comparing the top-hat model and the two-component model, the curvature-tail effect starting times are the same, given the parameters above, and both of them have the temporal decay indices steeper than that of the spherical model in the black dotted curve. However, the two-component model has a shallower decay after the curvature-tail effect starts due to the contribution of the second component. This new contribution makes temporal decay indices tend to get back on their original track after $t_{\rm obs,2}$. Therefore, there is a sharp drop in the lower panel. Then, the cut-off occurs due to the curvature-tail effect on the second component.

The jet models considered in the above discussions are characterized by sharp and discontinuous changes in their properties at the jet edges. Such abrupt boundaries may be physically unrealistic. To explore more plausible scenarios, we now introduce two structured jet models that feature a smoother transition from the core to the surrounding wing and medium: (1) a jet with a power-law wing and (2) a jet with a Gaussian wing. In both models, the core remains uniform, but the energy injected into the wings follows a continuous functional form. We will define these two models and illustrate their characteristic light curves with the canonical physical parameters.

The power-law wing structure is assumed to have a narrow core with uniformly distributed injection for $\theta < \theta_{\rm core}$. The external component has a power-law injection distribution for $\theta_{\rm core} <\theta < \theta_{\rm w}$. The injection term in the electron continuity equation can be expressed as:
\begin{equation} \label{eq:injection_powerlaw}
Q'\left( \gamma^{\prime}_{\rm e}\right)=
\begin{cases}
Q_{\rm 0} {\left( \dfrac{\gamma^{\prime}_{\rm e}}{\gamma^{\prime}_{\rm m} }\right)}^{-p}, & 0 < \theta < \theta_{\rm core}\\
Q_{\rm 0} {\left( \dfrac{\theta}{\theta_{\rm core}} \right)}^{-l}{\left( \dfrac{\gamma^{\prime}_{\rm e}}{\gamma^{\prime}_{\rm m} }\right)}^{-p}, & \theta_{\rm core} < \theta < \theta_{\rm w}
\end{cases},
\end{equation}
where $l$ is the angular power-law index of the electron injection distribution. In this subsection, we fix $l=3$. 

Since the difference between the two conditions in Equation \eqref{eq:injection_powerlaw} is just the magnitude, the way we achieve it numerically is simply adding a coefficient to finitely angle-divided sections of the entire wing. To test the power-law wing model, we set $\theta_{\rm core}=0.02 ~ {\rm rad}$ and $\Delta\theta_{\rm wing}=0.01 ~ {\rm rad}$ and calculate the light curves. The power-law wing model is plotted in purple in Figure \ref{fig:model_gradually_two}. Compared to the uniform wing model, the power-law wing model has a continuous evolution of the decay index and tends to have steeper decay indices after the wing component starts to dominate, which is in line with our expectations.

As for the wing with a Gaussian structure, the electron injection term is set as:
\begin{eqnarray} \label{eq:injection_Gaussian}
Q'\left( \gamma^{\prime}_{\rm e}\right)=&&
\begin{cases}
Q_{\rm 0} {\left( \dfrac{\gamma^{\prime}_{\rm e}}{\gamma^{\prime}_{\rm m} }\right)}^{-p}, ~~~~~~~~~~~ 0 < \theta < \theta_{\rm core}\\
Q_{\rm 0} \exp{\left(-\dfrac{1}{2} \dfrac{(\theta-\theta_{\rm core})^2}{\theta_0^2} \right)}{\left( \dfrac{\gamma^{\prime}_{\rm e}}{\gamma^{\prime}_{\rm m} }\right)}^{-p}, 
\end{cases}
\nonumber\\
&& ~~~~~~~~~~~~~~~~~~~~~~~~~~~~~~~\theta_{\rm core} < \theta < \theta_{\rm w},
\end{eqnarray}
where $\theta_0$ is a free parameter and is artificially set as 0.007 rad for the calculation in this subsection. By this definition, there is a smooth transition at the core-wing interface. To test the jet model with a Gaussian wing, we set similar angles as those in the power-law wing model, where $\theta_{\rm core}=0.02 ~{\rm rad}$ and $\Delta\theta_{\rm wing}=0.01~{\rm rad}$. The Gaussian wing model is plotted in pink in Figure \ref{fig:model_gradually_two}. The properties of the curvature-tail effect in the Gaussian wing are similar to those shown in the power-law wing. However, with a shallower decay index after the Gaussian wing starts to dominate, the decay index has a significantly faster increasing speed, and the light curve has an earlier cut-off.

\section{Fit to GRB 230307A} \label{sec:TheFit}
As established in the above sections, the detection of a curvature-tail effect during the prompt emission is observationally challenging, requiring an event that has both exceptionally high luminosity and narrowly collimated jet. The recent burst GRB 230307A meets these criteria exceptionally well and provides an opportunity to test this model. As one of the brightest GRBs ever detected, it readily satisfies the luminosity requirement. Crucially, an analysis of data from the Gravitational wave high-energy Electromagnetic Counterpart All-sky Monitor (GECAM) \citep{2022RDTM....6...12L,2023NIMPA105668586Z} by \cite{2025NSRev..12E.401S} revealed a distinct steepening in the late-time light curve of prompt emission, which is the precise signature predicted by the curvature-tail effect. These unique characteristics make GRB 230307A an ideal case for our curvature-tail effect model.

\begin{figure}[ht!]
\centering
\includegraphics[width=1.01\linewidth]{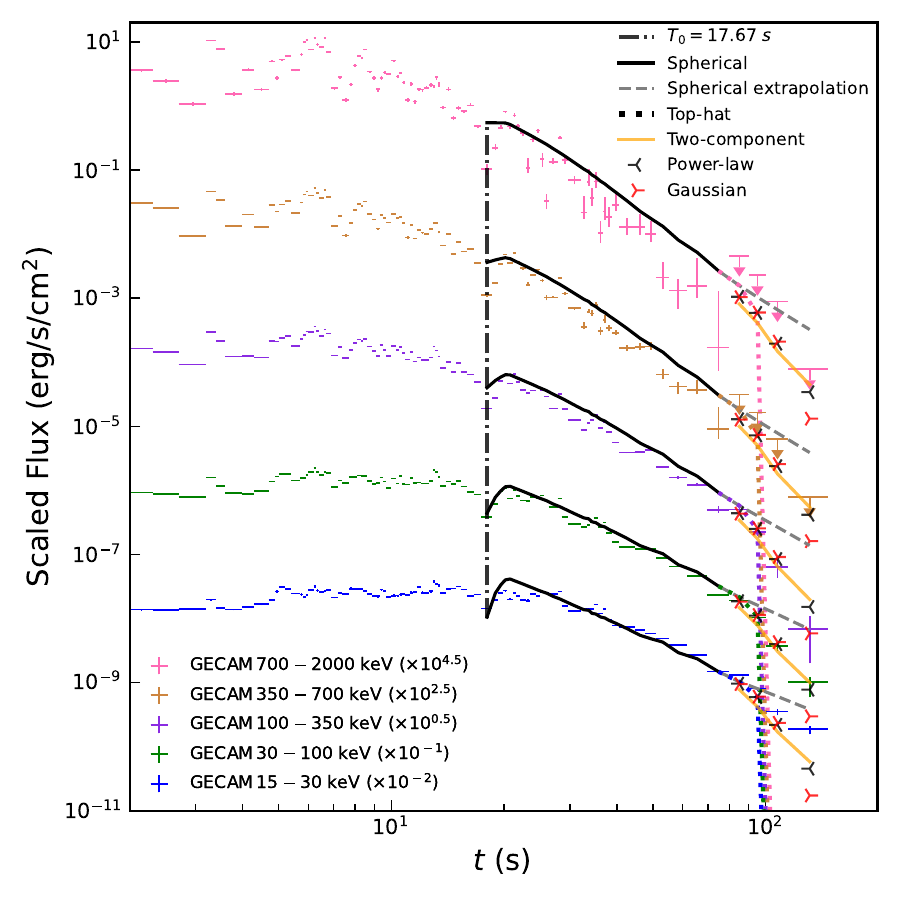}
\caption{Scaled flux of the light curves on a log-log scale predicted by the models compared with GRB 230307A's late-time light curves using GECAM data. The points with error bars and upper limits are from observed GECAM data. The black solid curves within 18 s to 84 s represent the prediction of the best-fit parameters of the spherical model without the curvature-tail effect. The grey dashed curves after 84 s are the extrapolation of the spherical model. The colored dotted curves, the orange solid curves, the black tri-left markers, and the red tri-right markers represent the best-fit models with the top-hat jet, the two-component jet, the power-law wing, and the Gaussian wing, respectively. The black dash-dotted vertical line represents the time when the fit starts.
The error bars represent the $1\sigma$ uncertainties. 
\label{fig:fit}}
\end{figure}

In this work, we use the synchrotron model in Equation \eqref{eq:Fv_total} and \eqref{eq:lower time limit} with different jet structures to fit the late-time light curve of GRB 230307A. The fitting uses the Python module {\it emcee} \citep{2013PASP..125..306F} to perform the Bayesian parameter estimation, and the log-likelihood function can be expressed as
\begin{equation} \label{eq:likelihood}
 \ln \mathcal{L}=-\frac{1}{2}\sum_i \Big[\ln (2\pi\sigma_i^2)+\frac{(y_i-m_i)^2}{\sigma_i^2}\Big],
\end{equation}
where $y_i$ and $\sigma_i$ are the flux and error of GECAM data and $m_i$ is the model-predicted flux. The GECAM data analysis procedure is the same as \cite{2025NSRev..12E.401S}. We used the analyzed light curve data from five energy bands of GECAM, 15--30, 30--100, 100--350, 350--700, 700--2000 keV, and jointly fit them with a single set of parameters. The observed light curves with errors and upper limits in five energy bands are present in Figure \ref{fig:fit}. 

The selection of the temporal window for our model fit requires careful consideration. Our primary goal is to model the whole curvature effect phase that includes the late-time steepening observed in GRB 230307A, where the beginning of the curvature effect phase is around $t_{\rm obs}\sim$ 23 s, the average ``$t_{\rm b2}$" of the five energy bands in ``Supplementary Data Table 4" of \cite{2025NSRev..12E.401S}. A physically complete model should also encompass the pulse's rising phase. However, the light curve before this time is complex, featuring multiple overlapping pulses and a prominent achromatic dip around 18 s \citep{2025ApJ...985..239Y}. Attempting to fit this entire complex light curve with a single-pulse model would be inappropriate and yield a poor fit. Fortunately, the emission following the dip at around 18 s presents a relatively clean and well-defined broad pulse structure with both a rising and a long decaying phase, even if the ongoing small-scale variability in the light curve remains outside of the scope of our model. We therefore make the simplifying assumption that the light curve after this dip can be modeled as a single emission event. 

The initial portion of the second broad pulse of GRB 230307A after the dip is buried under the tail of the first pulse before the dip. Therefore, the real zero time of the second broad pulse should be $T_0=T_0^{\rm obs}-(\Delta T)^{\rm offset}$, where $T_0^{\rm obs}$ is the apparent zero time in observation and $(\Delta T)^{\rm offset}$ is the zero-time offset missed and can not be straightforwardly identified from the data. Therefore, at time $t_{\rm obs}$, the duration of the second pulse is $t_{\rm obs}-T_0$. In our initial attempts, we find that adding a new free $(\Delta T)^{\rm offset}$ in the synchrotron model makes the other fitting parameters difficult to constrain. Therefore, in our final fitting method, $(\Delta T)^{\rm offset}$ is determined independently from 100--350 keV (middle energy band) light curve fitting using an empirical model, $I(t_{\rm obs})=A {\rm exp}[2\mu-\tau_1/(t_{\rm obs}-T_0)-(t_{\rm obs}-T_0)/\tau_2]=A {\rm exp}\{2\mu-\tau_1/[t_{\rm obs}-T_0^{\rm obs}+(\Delta T)^{\rm offset}]-[t_{\rm obs}-T_0^{\rm obs}+(\Delta T)^{\rm offset}]/\tau_2\}$ \citep{2005ApJ...627..324N}. We fix $T_0^{\rm obs}=18~{\rm s}$ and obtain $(\Delta T)^{\rm offset}=0.33_{-0.02}^{+0.02} ~{\rm s}$. Therefore, $T_0=T_0^{\rm obs}-(\Delta T)^{\rm offset}=17.67 ~{\rm s}$ is used in synchrotron model fitting.

We then fit our model to the GECAM light curve in all five energy bands jointly across the time interval from 18 s ($T_0^{\rm obs}$) to 131 s. This approach allows us to model the complete final pulse while isolating it from the complex earlier emission. However, our initial attempts to fit the entire post-dip light curve from 18 s to 131 s at once were unsuccessful. The data after the break at around 84 s have significantly higher uncertainties and fewer data points than the pre-break data. Consequently, the low statistical weight of this late-time segment meant that the likelihood function in Equation \eqref{eq:likelihood} was insensitive to the post-break data, and the parameters governing the jet structure failed to converge. To overcome this challenge, we adopted a two-step fitting strategy. First, we fit the high-quality, pre-break data from 18 s to 84 s using an isotropic spherical model. The 84 s above is around the average ``$t_{\rm b3}$" in ``Supplementary Data Table 4" of \cite{2025NSRev..12E.401S}, which indicates the time when the late-time break occurs. This model is appropriate for the early phase, where the jet edge is not yet visible, and allows us to constrain the physical parameters unrelated to the jet geometry. Second, we fixed these best-fit parameters and then fit our structured jet models only to the post-break data from 84 s to 131 s, with the jet geometry parameters ($\theta_{\rm jet}$, etc.) as the only free variables. This approach ensures that the pre-break light curve remains consistent across all jet models as long as the jet has a uniform core, while allowing the statistically weaker post-break data to exclusively determine the jet structure.

In our initial fitting attempts, the decay index of magnetic field strength $\alpha_B$ of the spherical model in Equation \eqref{eq:Fv_total} is a free parameter, but we find that it is difficult to constrain.
Then we tried to adjust the value of $\alpha_B$ from 0 to 2 and fix all the other best-fit parameters to check how light curves change. The results show that $\alpha_B$ does not have a significant influence on the light curve of the second broad pulse of this burst. To reduce the number of free parameters, we fix $\alpha_B$ to 0 in our final fitting.

The fitting results of the isotropic spherical jet model and a top-hat jet model are presented in Figure \ref{fig:fit}, and the corner plot is shown in Figure \ref{fig:corner_modelA1} in Appendix \ref{sec:cornerplot}. The best-fit parameters are defined as the ones with the highest likelihood within their $1\sigma$ regions. The best-fit parameters and fitting statistics are listed in Table \ref{table:fit} and their prior ranges of the fits in Table \ref{table:prior}. The fits have high $\chi^2$ statistics because the high variability in the light curve cannot be captured in the model.

\begin{table*} 
\centering
\addtolength{\leftskip} {-2.3cm}
\caption{Best-fit Parameters of Jet Models}
\label{table:fit}
\resizebox{1.1\textwidth}{!}{
\begin{threeparttable}
\begin{tabular}{lcccccccc}
\toprule
Pre-break Model & $\mathrm{log} [B^{\prime}_{\rm 0}(\mathrm{G})]$ & $\mathrm{log} \gamma^{\prime}_{\rm m}$ & $\mathrm{log} \Gamma$ & $p$ & $t_\mathrm{inj}(\mathrm{s})$ & $\mathrm{log} [R_\mathrm{0}(\mathrm{cm})]$ & $\mathrm{log} [Q_0({\rm cm}^{-1})] $ & $\chi^2$/dof \\ 
\hline
Spherical & $2.79_{-0.02} ^{+0.03}$ & $3.98_{-0.01}^{+0.04}$ & $2.10_{-0.03} ^{+0.02}$ & $3.45_{-0.07} ^{+0.18}$ & $2.68_{-0.05} ^{+0.10}$ & $16.24_{-0.05} ^{+0.02}$ & $32.40_{-0.05} ^{+0.07}$ &43849.59/152 \\
\hline
\hline
Post-break Model & $\theta_{\rm jet}$ & $\theta_{\rm core}$ & $\Delta\theta_{\rm wing}$ & $f$ & $l$ & $\theta_0$ & $\chi^2$/dof & \\ 
\hline
Top-hat & $0.0156_{-2.9 \times 10^{-4}} ^{+2.9 \times 10^{-6}}$ & -- & -- & -- & -- & -- & 211.77/11 & \\
Two-component & -- & $0.0151_{-1.5 \times 10^{-2}} ^{+2.0 \times 10^{-4}}$ & $0.0031_{-3.1 \times 10^{-4}} ^{+5.9 \times 10^{-3}}$ & $0.385_{-4.0 \times 10^{-2}} ^{+3.9 \times 10^{-2}}$ & -- & -- &237.36/9 & \\
Power-law Wing & -- & $0.0147_{-8.3 \times 10^{-5}} ^{+5.7 \times 10^{-5}}$ & $0.0132_{-6.4 \times 10^{-3}} ^{+4.7 \times 10^{-3}}$ & -- & $9.82_{-7.9 \times 10^{-1}} ^{+1.2 \times 10^{-1}}$ & -- &85.45/9 & \\
Gaussian Wing & -- & $0.0140_{-2.1 \times 10^{-4}} ^{+1.7 \times 10^{-4}}$ & $0.0067_{-2.3 \times 10^{-3}} ^{+9.0 \times 10^{-3}}$ & -- & -- & $0.0017_{-1.6 \times 10^{-4}} ^{+2.0 \times 10^{-4}}$ &97.72/9 & \\
\hline
\hline
\end{tabular}
\begin{tablenotes}
\addtolength{\leftskip} {2.25cm}
\footnotesize
\item {\it Note}. The error bars represent the $1\sigma$ uncertainties.
\end{tablenotes}
\end{threeparttable}}
\end{table*}

\begin{table*} 
\addtolength{\leftskip} {0.8cm}
\caption{Prior Ranges of the Fitting Parameters}
\label{table:prior}
\resizebox{0.76\textwidth}{!}{
\begin{threeparttable}
\begin{tabular}{lccccccc}
\toprule
Pre-break Model & $\mathrm{log} [B^{\prime}_{\rm 0}(\mathrm{G})]$ & $\mathrm{log} \gamma^{\prime}_{\rm m}$ & $\mathrm{log} \Gamma$ & $p$ & $t_\mathrm{inj}(\mathrm{s})$ & $\mathrm{log} [R_\mathrm{0}(\mathrm{cm})]$ & $\mathrm{log} [Q_0({\rm cm}^{-1})] $ \\ 
\hline
Spherical & [-1,4] & [3.5,6] & [2,4] & [2,5] & [0,5] & [13,16.5] & [28,34] \\
\hline
\hline
Post-break Model & $\theta_{\rm jet}$ & $\theta_{\rm core}$ & $\Delta\theta_{\rm wing}$ & $f$ & $l$ & $\theta_0$ & \\ 
\hline
Top-hat & [0,0.05] & -- & -- & -- & -- & -- & \\
Two-component & -- & [0,1] & [0,1] & [0.1,10] & -- & -- & \\
Power-law Wing & -- & [0,0.02] & [0,0.02] & -- & [2,10] & -- & \\
Gaussian Wing & -- & [0,0.02] & [0,0.02] & -- & -- & [0,0.02] & \\
\hline
\hline
\end{tabular}
\begin{tablenotes}
\addtolength{\leftskip} {2.05cm}
\footnotesize
\item {\it Note}. 
The error bars represent the $1\sigma$ uncertainties.
\end{tablenotes}
\end{threeparttable}}
\end{table*}

In Figure \ref{fig:fit}, we present the comparison between the late-time light curves of GECAM data and model predictions with the isotropic spherical jet model and a top-hat jet model from 18 s to 131 s. The solid black curves represent the prediction of the spherical model and fit the observation well before the break. However, its extrapolation in the dashed grey curve after the break significantly overestimates the observation and exceeds the upper limits in the high-energy bands. This means the jet has a more complex structure, and the structure provides observable signatures around 84 s. Then we use the top-hat model to fit the part after 84 s. In this model, there is only one free parameter, the jet half-opening angle $\theta_{\rm jet}$, and all the other parameters are fixed as the best-fit values of the spherical model. 

A notable result from our analysis is the consistency between the best-fit half-opening angle for the top-hat jet, $\theta_{\rm jet} = 0.0156$ rad, and the theoretically expected angle. This theoretical angle can be obtained from Equation \eqref{eq:theta_max} assuming the break time $t_{\rm break}=84 \ {\rm s} -17.67 \ {\rm s}$:
\begin{equation} \label{eq:DerivedAngle}
\theta_{\rm jet}^{*}=\arccos{\left( 1-\dfrac{ct_{\rm break}}{(1+z)R_0} \right)}=0.0147 ~{\rm rad}.
\end{equation} 
Their relative error is $\delta=(\theta_{\rm jet}-\theta_{\rm jet}^{*})/\theta_{\rm jet}^{*}=6.12\%$. This consistency between the detailed numerical fit and the analytical prediction provides strong validation for our modeling framework and fitting procedure.

From the model-predicted light curves of the top-hat jet model in the colored dotted curves, a cut-off occurs prominently. As a result, the model-predicted light curves have much steeper temporal decay indices after the break and underestimate the observation, compared with those of GECAM data in the three lower energy bands (15--30, 30--100, and 100--350 keV). Our results demonstrate that a simple top-hat jet model still can not reproduce the observed late-time light curve of GRB 230307A. This failure suggests that the jet is not sharply truncated but may possess additional emission components outside the primary internal core, such as a structured ``wing" as proposed by \cite{2025NSRev..12E.401S}. The presence of such a structure would produce a shallower decay after the break compared with a top-hat jet. This motivates the investigation to perform a detailed fit of each wing model to the GRB 230307A data to determine which, if any, can provide a self-consistent explanation for the observed emission.

Therefore, we fit the three models in Section \ref{sec:wing structure model} to the light curves after 84 s. The best-fit parameters are listed in Table \ref{table:fit}, their prior ranges of the fits in Table \ref{table:prior}, and the model-predicted light curves can be found in Figure \ref{fig:fit}.

As shown in Figure \ref{fig:fit}, the best-fit light curves for all three structured jet models, the two-component jet model, the power-law wing model, and the Gaussian wing model are similar and provide a better description of the data than that of the top-hat model. The presence of the wing component results in a higher post-break flux, producing a shallower decay that is consistent with the observations. Importantly, all three models retain the core curvature-tail break signature, exhibiting a clear steepening after 84 s. It is worth mentioning that the wing models still slightly underestimate a few data points in the lowest energy band, 15--30 keV. This may result from the simplified assumptions in the jet models and magnetic field evolution, while the actual situation can be much more complicated.

A comparison of reduced $\chi^2$ values among the three jet models with wings indicates that the power-law wing jet model is favored. The best-fitting configuration consists of a uniform core ($\theta_{\rm core}=0.0147$ rad) and a uniform wing ($\Delta\theta_{\rm wing}=0.0132$ rad), providing a good description of the observed light-curve break of GRB 230307A during the prompt emission phase.

The two-step fitting strategy has been proven successful with the provided data. Although the fit presented in this section exclusively utilizes data collected after 84 s in three low-energy bands, the continuity and the smoothness with the spherical model around 84 s are preserved in the new models. This is due to the constraint of fixing the eight parameters previously obtained from the spherical model. Therefore, all models consistently depict the same temporal evolution and exhibit identical magnitudes before the curvature-tail effect occurs.

\section{Discussion} \label{sec:Discussion}

In this work, we propose the curvature-tail effect and quantitatively examined this geometric effect within the curvature-emission framework using the prompt emission of GRB 230307A. A theoretical achromatic break in the light curve naturally emerges when the jet edge comes into view, which is consistent with the observations of GRB 230307A. Furthermore, adopting a structured jet model rather than a simple top-hat jet provides a better description of the post-break behavior, demonstrating that the signatures of curvature-tail effect in the prompt phase are a powerful diagnostic of the angular structure of GRB jets.

Despite its diagnostic potential, the curvature-tail effect is observationally challenging to detect, as several conditions must be satisfied. The burst must be sufficiently bright, and its decay phase needs to be monitored long enough to capture the break with high significance. A small jet opening angle is also favorable, since in this case the break occurs before the flux has declined substantially, making the feature easier to detect. GRBs often exhibit complex, multi-pulse structures. Although overlapping pulses can obscure the signal, the decay of the final pulse should still carry the expected break if the temporal coverage is adequate. Furthermore, because high-latitude emission is Doppler-shifted toward lower energies, broadband coverage, particularly extending into the X-ray band, is crucial for identifying this feature. These requirements account for the rarity of observed curvature-tail breaks while also indicating the observational strategies that are most effective for detecting them. 

Our study of GRB 230307A demonstrates that curvature-tail signatures from the prompt phase can be directly observed and modeled, providing new constraints on the jet’s angular structure. In the absence of afterglow jet break observations, such prompt-phase signatures can place an independent constraint on jet structure. When afterglow data provide sufficiently strong constraints on the jet structure (e.g. \citealt{ghirlandaCompactRadioEmission2019}), the two approaches can be combined to offer complementary constraints and powerful cross-validation, leading to a more reliable understanding of GRB jet geometry. A promising direction for future work is to extend this analysis to a larger sample of bright and long GRBs with broadband coverage, where curvature-tail features may be systematically identified. For instance, GRB 221009A might be a potential candidate due to its being the brightest GRB, and its narrow jet half-opening angle, $0.8^\circ$ \citep{2023Sci...380.1390L}. Additionally, GRB 240825A can be another potential candidate, because \cite{2025ApJ...985L..30W} estimates its jet opening angle using the achromatic break of prompt emission in the light curve, and they suggest a possible two-component jet for the burst.

The best-fit parameters of GRB 230307A provide additional clues to the underlying physics and enable a direct comparison with previous work. The best-fit value of the core angle of the models, as shown in Table \ref{table:fit} is around 0.015 rad while the angle estimated in \cite{2025NSRev..12E.401S} is $3.4 ^{\circ} \sim 0.06$ rad. One possible reason is that \cite{2025NSRev..12E.401S} assumed an emission radius $R_0 = 10^{15}$ cm, while the radius is a free parameter in the fit in this work. The best-fit radius is $10^{16.24}$ cm, and this can lead to the difference. Although such a large radius is unusual, it remains possible within some dissipation models (e.g., \citealt{2009MNRAS.395..472K,2025JHEAp..4700359Y,2025ApJ...985..239Y}). The opening angle constrained in this work is also somewhat smaller than that inferred from afterglow fitting (e.g., $\log \theta_{\rm c}=-1.38_{-0.27}^{+0.29}$, \citealt{2024Natur.626..742Y}). Since our constraint is based on the prompt emission, the difference may arise from lateral expansion of the jet during its evolution toward the afterglow phase, and it should be noted that narrower jets are expected to experience more rapid sideways spreading \citep{granotLateralExpansionGammaray2012a}.

This work investigates how jet-structure information can be extracted from light-curve breaks in the prompt emission, using the curvature-tail effect. We note that the morphology of such breaks is not solely a function of the angular energy profile; rather, it depends on at least three additional critical factors: (i) the intrinsic prompt spectrum and its spectral evolution governed by the underlying mechanism (e.g., internal shock model \citep{1994ApJ...427..708P,1994ApJ...430L..93R,1998MNRAS.296..275D}, ICMART model \citep{2011ApJ...726...90Z} or photosphere model \citep{1991ApJ...369..175A,2000ApJ...530..292M}), which dictates the contribution of high-latitude emission to the observed band; (ii) the geometry of the prompt cessation surface and its mapping via the EATS relation; and (iii) the definition of the zero-time reference $T_0$, which may significantly influence the inferred decay indices and fitted physical parameters. 
The following briefly summarizes several physically motivated assumptions that have been clarified in Section \ref{sec:TheFit} for these factors. The emission is modeled as a synchrotron process with adiabatic cooling, initiated at a free radius $R_0$. Moreover, our framework assumes a spherical cessation surface confined within the jet cone, characterized by a uniform Lorentz factor $\Gamma$ profile. 
Furthermore, $T_0=T_0^{\rm obs}-(\Delta T)^{\rm offset}=17.67 ~{\rm s}$, which is obtained from empirical model fitting, is used in synchrotron model fitting. Future studies incorporating alternative energy dissipation mechanisms and radiation mechanisms, non-uniform $\Gamma$ profiles, and a free zero-time offset $(\Delta T)^{\rm offset}$ in the physical model will be essential to further break the inherent degeneracies in prompt emission modeling.

It is worth noting that the curvature-tail effect discussed in this paper becomes significant when the magnetic field is relatively strong, so that the electrons responsible for the observed emission are nearly all in the fast-cooling regime. In this case, the resulting light-curve break appears nearly achromatic across different energy bands, as the emission declines rapidly once the electrons cool. However, if the magnetic field is weaker or decays more rapidly with time, the radiative cooling time becomes longer, and the light-curve evolution is affected by the combined action of the cooling process and the curvature-tail effect. As a result, the break caused by the curvature-tail effect becomes less distinct, and the effect itself becomes energy dependent: light curves in the high-energy bands, produced mainly by high-energy electrons, are still dominated by the curvature-tail effect, whereas those in the low-energy bands, arising from lower-energy electrons, are governed by a combination of slow-cooling emission and geometric effects. Consequently, the curvature-tail effect produces a chromatic break, with the high-energy bands exhibiting an earlier and steeper decay than the lower-energy ones. The interplay between the cooling regime and the curvature-tail effect can therefore introduce considerable complexity into the temporal evolution of the prompt light curves, which will be examined in future work.

\section*{Acknowledgement}
We are grateful to the referee for helpful comments. We thank Yi-Han Iris Yin, Rudra D Patel, and Yue Wu for helpful comments. We acknowledge the support by the National Natural Science Foundation of China (grants 12573046 and 12121003 to Bin-Bin Zhang and grants 12393811 and 12473048 to Xiao-Hong Zhao), the science research grants from the China Manned Space Project (grant CMS-CSST-2021-B11 to Bin-Bin Zhang). Bin-Bin Zhang acknowledges support by the Fundamental Research Funds for the Central Universities, and the Programme for Innovative Talents and Entrepreneurs in Jiangsu. Bin-Bin Zhang and Hendrik van Eerten acknowledge support by Royal Society International Exchanges Cost Share 2022 NSFC award IEC$\backslash$NSFC$\backslash$223149. Zhen-Yu Yan acknowledges support by China Scholarship Council (202406190271) and the University of Bath Visiting Postgraduate Scholar program (249552753). Zhen-Yu Yan is also supported by the Program for Outstanding PhD Candidates of Nanjing University (2025A06). 

\bibliography{main}
\bibliographystyle{aasjournal}

\appendix
\counterwithin{figure}{section}
\counterwithin{table}{section}
\counterwithin{equation}{section}

\section{The definitions of the key times} \label{sec:time}
\counterwithin{figure}{section}
\counterwithin{table}{section}
\counterwithin{equation}{section}

\begin{table}[h] 
\centering
\addtolength{\leftskip} {-0.3cm}
\caption{The definitions of the key times}
\label{table:time}
\resizebox{1.\textwidth}{!}{
\begin{threeparttable}
\begin{tabular}{lcr}
\toprule
Name & Definition & Reference frame \\ 
\hline
$t_{\rm obs}$ & the time after the trigger of the burst & observer frame \\
$t_{\rm obs,0}$ & the duration of a shell traveling from central engine to emission region & observer frame \\
$t_{\rm obs,1}$ & \makecell[l]{the time when the maximum angle on the EATS in emission region equals the jet's half-opening angle \\(the time when one observes the first photon from the newly injected electron on the jet edge)} & observer frame \\
$t_{\rm obs,2}$ & \makecell[l]{the time when the EATS touches the jet edge at injection stopping radius of electron \\(the time when one observes the first photon from the last injected electron on the jet edge)} & observer frame \\
$t_{\rm inj}$ & the total injection duration of electron & observer frame \\
$t_{\rm e}$ & the time after the shell leaves central engine & source frame \\
$T_{\rm 0}^{\rm obs}$ & the apparent zero time of the second broad pulse of GRB 230307A & observer frame \\
$(\Delta T)^{\rm offset}$ & the zero-time offset between the apparent zero time and the real zero time of the second pulse of GRB 230307A & observer frame \\
\hline
\hline
\end{tabular}
\end{threeparttable}}
\end{table}

\section{The corner plot of the best-fit results using the spherical model} \label{sec:cornerplot}
\counterwithin{figure}{section}
\counterwithin{table}{section}
\counterwithin{equation}{section}

\begin{figure*}[htbp]
\centering
\includegraphics[width=0.69\linewidth]{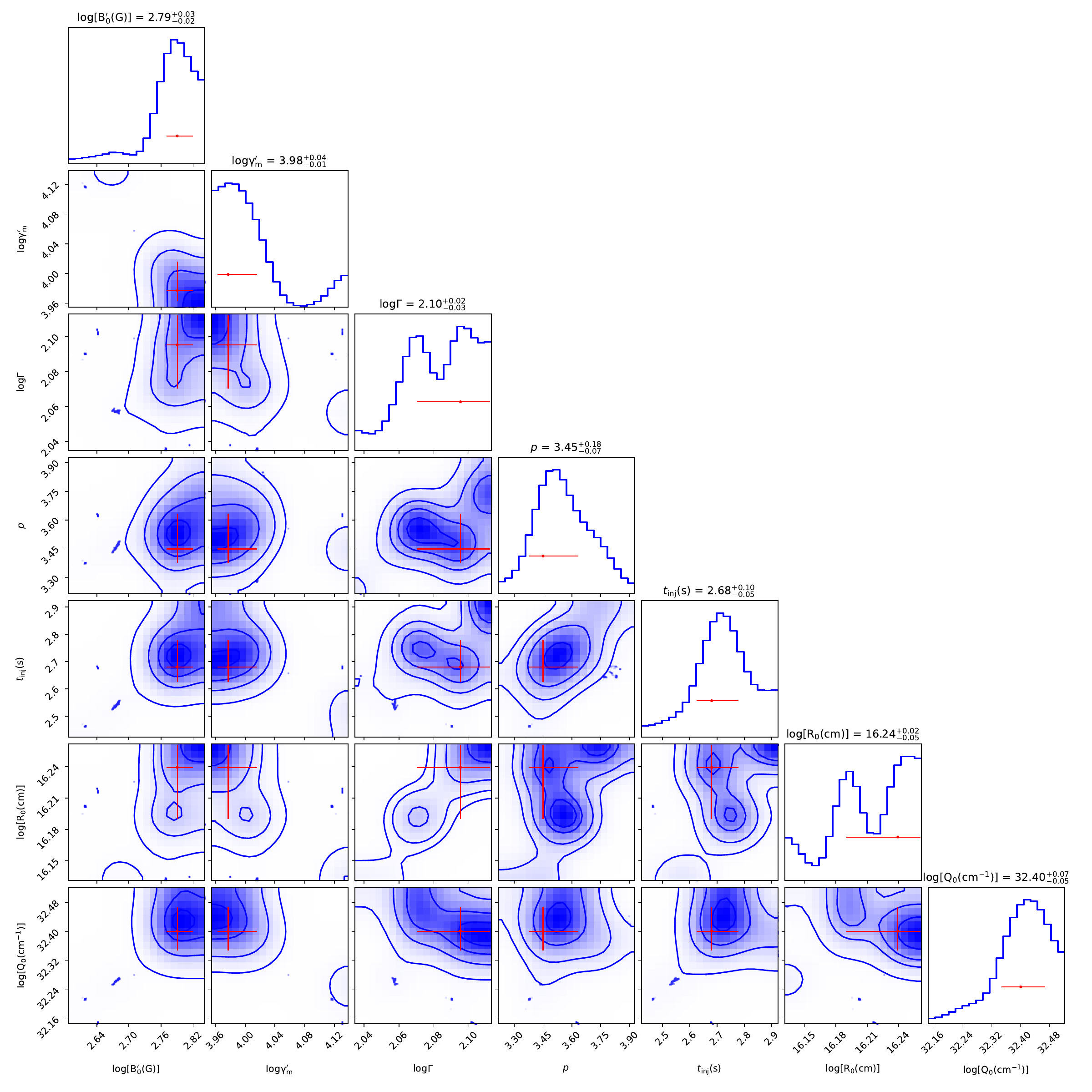}
\caption{Corner plot of the best fit of the isotropic spherical model to the late-time light curve of GRB 230307A with GECAM data. The error bars represent the $1\sigma$ uncertainties. 
\label{fig:corner_modelA1}}
\end{figure*}

\end{document}